\documentclass[11pt,a4paper]{article}

\usepackage{epsfig,amsmath,amssymb,cite}

\begin{document}

\title{Thin--shell wormholes with charge in $F(R)$ gravity} 
\author{Ernesto F. Eiroa$^{1, 2}$\thanks{e-mail: eiroa@iafe.uba.ar}, Griselda Figueroa Aguirre$^{1}$\thanks{e-mail: gfigueroa@iafe.uba.ar}\\
{\small $^1$ Instituto de Astronom\'{\i}a y F\'{\i}sica del Espacio (IAFE, CONICET-UBA),}\\
{\small Casilla de Correo 67, Sucursal 28, 1428, Buenos Aires, Argentina}\\
{\small $^2$ Departamento de F\'{\i}sica, Facultad de Ciencias Exactas y 
Naturales,} \\ 
{\small Universidad de Buenos Aires, Ciudad Universitaria Pabell\'on I, 1428, 
Buenos Aires, Argentina}} 

\maketitle

\begin{abstract}
In this article, we construct a class of constant curvature and spherically symmetric thin--shell Lorentzian wormholes in $F(R)$ theories of gravity and we analyze their stability under perturbations preserving the symmetry.  We find that the junction conditions determine the equation of state of the matter at the throat. As a particular case, we consider configurations with mass and charge. We obtain that stable static solutions are possible for suitable values of the parameters of the model.\\

\noindent 
PACS number(s): 04.20.Gz, 04.40.Nr, 98.80.Jk\\
Keywords: Lorentzian wormholes; exotic matter; $F(R)$ theories

\end{abstract}

\section{Introduction}\label{intro} 

Traversable Lorentzian wormholes are solutions of gravitational theories which have a throat that connects two regions of the same universe or two different universes \cite{motho,visser}. In General Relativity, they are threaded by matter that violates the null energy condition \cite{motho,visser,hovis}; the amount of this exotic matter can be made arbitrary small \cite{viskardad}, but at the expense of large pressures at the throat \cite{lst}. Traversable wormholes can be constructed \cite{visser} by cutting and pasting two manifolds to form a new one, with a shell at the joining surface corresponding to the throat, where the flare--out condition is fulfilled. These thin--shell wormholes have been extensively studied in the literature because of their simplicity, which makes the analysis of stability easier, and the exotic matter can be confined to the shell. Wormholes with a continuous energy-stress tensor at the throat usually also need a cut and paste procedure to confine the exotic matter or to obtain a suitable asymptotic behavior. Stability studies of spherically symmetric thin--shell wormholes, with a linearized equation of state at the throat, have been performed under radial perturbations by several authors (\cite{poisson,ishak-eirom-lobo,eir,dilem,lmv,shbiwh} and references therein). Plane and cylindrical thin--shell wormholes were also analyzed \cite{plane,dilem,cil}. The Chaplygin gas and its generalizations were used to model the exotic matter supporting wormholes \cite{chaplywh}. The linearized stability of Schwarzschild thin--shell wormholes with variable equations of state has been recently considered \cite{varela}.

Within the context of General Relativity, the observed accelerated expansion of the Universe during the matter dominated epoch requires of the presence of dark energy. Instead of this non--standard fluid, modifications of General Relativity were proposed in order to solve both the problems of dark energy and dark matter, required by the concordance ($\Lambda$CDM) model. One of the simplest possible modifications corresponds to the so called $F(R)$ gravity \cite{dft,sf,nojod}, in which the Einstein--Hilbert lagrangian is replaced by a function $F(R)$ of the Ricci scalar $R$. The $F(R)$ theories can provide an alternative for an unified picture of both inflation and the accelerated expansion at later times. Besides the cosmological aspects, it is of interest to study compact objects in these alternative theories. Static and spherically symmetric black hole solutions in $F(R)$ have been found \cite{bhfr1,bhfr2} in the last decade. Traversable wormholes in $F(R)$ were also studied in recent years \cite{whfr,bronnikov}.

Thin shells in General Relativity are modeled by using the well known Darmois--Israel \cite{daris} formalism. The junction conditions allow to match two solutions onto a hyper-surface under different conditions, for example the interior and exterior solutions corresponding to stars, galaxies, etc. They are also useful for the study of thin layers of matter and in braneworld cosmology.  In the last decade, the junction conditions have been generalized to $F(R)$ theories of gravity \cite{dss,js1}. The junction conditions are more stringent in $F(R)$ gravity than in General Relativity. For non--linear $F(R)$, they always require continuity of the trace of the second fundamental form at the matching hypersurface and, with the exception of quadratic $F(R)$, the continuity of the curvature scalar $R$. Quadratic $F(R)$ has some specific features: the curvature scalar $R$ can be discontinuous at the matching hypersurface and, as a consequence, the shell will have, besides the standard energy--momentum tensor, an external energy flux vector, an external scalar pressure (or tension) and another energy--momentum contribution resembling classical dipole distributions \cite{js2,js3}. The last one can be interpreted as a gravitational double layer. All these contributions should be present in order to make the whole energy--momentum tensor divergence--free \cite{js2,js3}. Recently, these results were extended to the  most general gravitational theory with a Lagrangian quadratic in the curvature \cite{js4}.

In the present work, we construct thin--shell wormholes with spherical symmetry in $F(R)$ theory with constant curvature and we study their stability under radial perturbations.  The paper is organized as follows: in Sect.  \ref{tswh}, the wormhole construction is done; in Sect. \ref{stab}, the stability of static configurations is analyzed; in Sect. \ref{charge} the formalism is applied to charged wormholes; finally, in Sect. \ref{conclu} the conclusions are presented. We adopt units in which $G=c=1$, where $G$ and $c$ denote the gravitational constant and the speed of light, respectively.

\section{Wormhole construction}\label{tswh}

We start from the spherically symmetric geometry defined by the metric 
\begin{equation} 
ds^2=-A (r) dt^2+A (r)^{-1} dr^2+r^2(d\theta^2 + \sin^2\theta d\phi^2),
\label{metric}
\end{equation}
where $r>0$ is the radial coordinate, $0\le \theta \le \pi$, and $0\le \varphi<2\pi $ are the angular coordinates. We adopt this metric in the construction of wormholes by using the thin--shell formalism in $F(R)$ gravity ($R$ is the curvature scalar). We choose a radius $a$ and we cut two identical copies of the region with $r\geq a$:
\begin{equation} 
\mathcal{M}^{\pm }=\{X^{\alpha }=(t,r,\theta,\varphi)/r\geq a\},  \label{e2}
\end{equation}
and paste them at the hypersurface
\begin{equation} 
\Sigma \equiv \Sigma ^{\pm }=\{X/G(r)=r-a=0\},  \label{e3}
\end{equation}
to create a new geodesically complete manifold $\mathcal{M}=\mathcal{M}^{+} \cup \mathcal{M}^{-}$. For a given $r$, the area $4\pi r^2$ is minimal when $r=a$, so the manifold  $\mathcal{M}$ represents a wormhole with two regions connected by a throat of radius $a$, where the flare--out condition is satisfied. A global radial coordinate can be defined on $\mathcal{M}$ by using the proper radial distance: $l=\pm \int_{a}^{r}\sqrt{1/A (r)}dr$, the signs $\pm $ correspond, respectively, to $\mathcal{M}^{+}$ and $\mathcal{M}^{-}$, and the throat is located in $l=0$. We denote the unit normals to $\Sigma $ in $\mathcal{M}$ by $n_{\gamma }^{\pm }$, the first fundamental form by $h_{\mu \nu}$, the second fundamental form  (extrinsic curvature) by $K_{\mu \nu}$, and the jump across the shell of any quantity $\Upsilon  $ by $[\Upsilon ]\equiv (\Upsilon ^{+}-\Upsilon  ^{-})|_\Sigma $.

Besides the continuity of the first fundamental form $h_{\mu \nu }$ across the shell, in  $F(R)$ theories there exists an additional condition \cite{js1}, which is $[K^{\mu}_{\;\; \mu}]=0$. If $F'''(R) \neq 0$ (the prime applied to $F(R)$ represents the derivative with respect to $R$), a third condition is also required \cite{js1}: the continuity of $R$ across the surface $\Sigma$, i.e.  $[R]=0$. However, quadratic $F(R)=R+\alpha R^2$, for which $F'''(R) = 0$, allows for the discontinuity of $R$. In our construction $[R]=0$ is automatically fulfilled, because the geometries at both sides of the throat are the same. The field equations on $\Sigma $ in the case $F'''(R) \neq 0$ have the form \cite{js1}
\begin{equation} 
\kappa S_{\mu \nu}=-F'(R)[K_{\mu \nu}]+ F''(R)[\eta^\gamma \nabla_\gamma R]  h_{\mu \nu}, \;\;\;\; n^{\mu}S_{\mu\nu}=0,
\label{LanczosGen}
\end{equation}
where $\kappa =8\pi $ and $S_{\mu \nu}$ represents the energy-momentum tensor at the shell. The field equations when $F'''(R) = 0$ read \cite{js1}
\begin{equation}
\kappa S_{\mu \nu} =-[K_{\mu\nu}]+2\alpha( [n^{\gamma }\nabla_{\gamma}R] h_{\mu\nu}-[RK_{\mu\nu}]), \;\;\;\; n^{\mu}S_{\mu\nu}=0,
\label{LanczosQuad}
\end{equation}
along with
\begin{equation}
\kappa\mathcal{T}_\mu=-2\alpha\nabla_\mu[R], \;\;\;\; n^{\mu}\mathcal{T}_\mu=0,
\end{equation}
\begin{equation}
\kappa\mathcal{T}=2\alpha [R] K^\gamma{}_\gamma ,
\end{equation}
\begin{equation}
\kappa \mathcal{T}_{\mu \nu}=2\alpha\Omega_{\mu \nu},
\end{equation}
where $\Omega_{\mu \nu}$ is a two-covariant symmetric tensor distribution 
\begin{equation}
\left<\Omega_{\mu\nu},\Psi \right> = -\int_\Sigma [R] h_{\mu\nu}  n^\gamma\nabla_\gamma \Psi,
\end{equation}
for any test function $\Psi$. In quadratic $F(R)$, besides the standard energy-momentum tensor $S_{\mu \nu}$, the shell has an external energy flux vector $\mathcal{T} _{\mu}$, an external scalar pressure/tension $\mathcal{T} $, and a double layer energy-momentum contribution $\mathcal{T}_{\mu \nu}$ of Dirac ``delta prime'' type, resembling classical dipole distributions  \cite{js1}. All of them are necessary in order to make the complete energy-momentum tensor divergence-free, so that it is locally conserved \cite{js1}. But all these contributions are proportional to $[R]$, so in our case they are all zero. Furthermore, for $[R]=0$, by using that $F'(R)=2\alpha R$ and $F''(R)=2\alpha$, it is also easy to see that  Eq. (\ref{LanczosQuad}) reduces to Eq. (\ref{LanczosGen}). The second term in the right hand side of Eq. (\ref{LanczosGen}) is zero because we are going to apply the formalism to geometries where $R=R_0$ is a constant. Then, in the case of $[R]=0$ and constant curvature $R=R_0$, we obtain that Eqs. (\ref{LanczosGen}) and (\ref{LanczosQuad}) both reduce to
\begin{equation}
\kappa S_{\hat{\imath}\hat{\jmath}}=-F'(R_0)[K_{\hat{\imath}\hat{\jmath}}],
\label{LanczosGen2}
\end{equation}
which is valid for any $F(R)$ theory.

At the surface $\Sigma $ we use the coordinates $\xi ^{i}=(\tau ,\theta,\varphi )$, with $\tau $ the proper time on the shell. We let the throat radius be a function of the proper time: $a(\tau)$.  The first fundamental form associated with the two sides of the shell is
\begin{equation}
h^{\pm}_{ij}= \left. g_{\mu\nu}\frac{\partial X^{\mu}}{\partial\xi^{i}}\frac{\partial X^{\nu}}{\partial\xi^{j}}\right| _{\Sigma },
\end{equation}
and the second fundamental form reads
\begin{equation}
K_{ij}^{\pm }=-n_{\gamma }^{\pm }\left. \left( \frac{\partial ^{2}X^{\gamma
} } {\partial \xi ^{i}\partial \xi ^{j}}+\Gamma _{\alpha \beta }^{\gamma }
\frac{ \partial X^{\alpha }}{\partial \xi ^{i}}\frac{\partial X^{\beta }}{
\partial \xi ^{j}}\right) \right| _{\Sigma },
\label{sff}
\end{equation}
with
\begin{equation}
n_{\gamma }^{\pm }=\pm \left\{ \left. \left| g^{\alpha \beta }\frac{\partial G}{\partial
X^{\alpha }}\frac{\partial G}{\partial X^{\beta }}\right| ^{-1/2}
\frac{\partial G}{\partial X^{\gamma }} \right\} \right| _{\Sigma }.
\end{equation}
For the metric (\ref{metric}) the unit normals ($n^{\gamma }n_{\gamma }=1$) take the form
\begin{equation}
n_{\gamma }^{\pm }=\pm \left(-\dot{a},\frac{\sqrt{A(a)+\dot{a}^2}}{A(a)},0,0 \right),
\end{equation}
where the dot represents the derivative with respect to $\tau$. Adopting the orthonormal basis $\{ e_{\hat{\tau}}=e_{\tau }, e_{\hat{\theta}}=a^{-1}e_{\theta }, e_{\hat{\varphi}}=(a\sin \theta )^{-1} e_{\varphi }\} $ at the shell, it is easy to obtain for the metric (\ref{metric}) that the first fundamental form is $h^{\pm}_{\hat{\imath}\hat{\jmath}}= \mathrm{diag}(-1,1,1)$, and the second fundamental form is given by
\begin{equation} 
K_{\hat{\theta}\hat{\theta}}^{\pm }=K_{\hat{\varphi}\hat{\varphi}}^{\pm
}=\pm \frac{1}{a}\sqrt{A (a) +\dot{a}^2}
\label{e4}
\end{equation}
and
\begin{equation} 
K_{\hat{\tau}\hat{\tau}}^{\pm }=\mp \frac{A '(a)+2\ddot{a}}{2\sqrt{A(a)+\dot{a}^2}},
\label{e5}
\end{equation}
with the prime on $A(r)$ representing the derivative with respect to $r$.  With the aid of Eqs. (\ref{e4}) and (\ref{e5}), the condition $[K^{\hat{\imath}}_{\;\; \hat{\imath}}]=0$ can be written in the form
\begin{equation} 
\ddot{a}=-\frac{A'(a)}{2}-\frac{2}{a}\left(A(a)+\dot{a}^2\right).
\label{CondGen}
\end{equation}
By replacing in Eq. (\ref{LanczosGen2}) the surface stress-energy tensor $S_{_{\hat{\imath}\hat{\jmath} }}={\rm diag}(\sigma , p_{\hat{\theta}}, p_{\hat{\varphi}})$, where $\sigma$ is the surface energy density and $p_{\hat{\theta}}$, $p_{\hat{\varphi}}$ are the transverse pressures, we find 
\begin{equation} 
\sigma= \frac{F'(R_0)}{\kappa\sqrt{A(a)+\dot{a}^2}}\left(2\ddot{a}+A'(a)\right)
\label{e9}
\end{equation}
and
\begin{equation}
p=\frac{-2 F'(R_0)}{\kappa a} \sqrt{A(a)+\dot{a}^2},
\label{e10}
\end{equation}
where $p=p_{\hat{\theta}}=p_{\hat{\varphi}}$. Replacing Eq. (\ref{CondGen}) in Eq. (\ref{e9}) we have
\begin{equation} 
\sigma= \frac{-4F'(R_0)}{\kappa a}\sqrt{A(a)+\dot{a}^2} ,
\label{e9cond}
\end{equation}
so if $F'(R_0)>0$ we can see that  $\sigma <0$, which indicates that the energy conditions are not satisfied, i.e. the matter at the throat is exotic. In this case the pressure is also negative, i.e. it is a tension. The inequality $F'(R)>0$ has an important interpretation in $F(R)$ gravity (see \cite{bhfr2} and references therein), because it implies that the effective Newton constant $G_{\mathrm{eff}}=G/F'(R)$ is positive. From the quantum point of view, $F'(R)>0$ prevents the graviton to be a ghost. So the energy conditions can be fulfilled only at the cost of the presence of ghosts. An interesting discussion about this topic in the case of wormholes can be found in Ref. \cite{bronnikov}. It is noteworthy that Eqs. (\ref{e10}) and (\ref{e9cond}) force an equation of state of the form 
\begin{equation}
p=\frac{\sigma}{2}.
\label{e11} 
\end{equation} 
By using this equation of state in combination with Eqs. (\ref{CondGen}), (\ref{e10}), and (\ref{e9cond}) it is easy to verify the conservation equation:
\begin{equation}
\frac{d(\sigma a^2)}{d\tau}+p\frac{d a^2}{d\tau}=0,
\label{conservacion}
\end{equation}
where the first term represents the internal energy change of the throat and the second one the work done by the internal forces of the throat.

\section{Stability of static configurations}\label{stab}

In the case of static wormholes with radius $a_0$, the condition given by Eq. (\ref{CondGen}) takes the form
\begin{equation} 
A'(a_0)=-4\frac{A(a_0)}{a_0},
\label{CondEstatico}
\end{equation}
from which we have
\begin{equation} 
a_0 A'(a_0)+4 A(a_0) = 0.
\label{CondEstatico-bis}
\end{equation}
The surface energy density and the pressure in the static case become, respectively,
\begin{equation} 
\sigma_0= \frac{F'(R_0)}{\kappa } \frac{A'(a_0)}{\sqrt{A(a_0)}}= \frac{-4 F'(R_0)}{\kappa a_0} \sqrt{A(a_0)},
\label{e13}
\end{equation}
and
\begin{equation}
p_0=\frac{-2F'(R_0)}{\kappa a_0} \sqrt{A(a_0)}.
\label{e14}
\end{equation}
For the study of the stability of static solutions under perturbations preserving the symmetry, we extend the method developed for General Relativity \cite{poisson} to $F(R)$ gravity. Using that $\ddot{a}= (1/2)d(\dot{a}^2)/da$ and defining $u=\dot{a}^2$, Eq. (\ref{CondGen}) can be written in the form 
\begin{equation}
u'(a)+\frac{4}{a}u=-A'(a)-\frac{4}{a}A(a).
\label{CondGen_u}
\end{equation}
By integrating Eq. (\ref{CondGen_u}) is possible to determine the dynamics of the throat,
\begin{equation}
\dot{a}^{2}=-V(a),
\label{condicionPot}
\end{equation}
where 
\begin{equation}
V(a)= A(a)-\frac{a_0^4}{a^4}A(a_0)
\label{potencial}
\end{equation}
can be interpreted as a potential. The same result can be found by introducing the equation of state (\ref{e11}) in the conservation equation (\ref{conservacion}) and integrating it to obtain $\sigma (a)=\sigma (a_0)a_0^3/a^3$; then replacing $\sigma (a)$ in Eq. (\ref{e9cond}) to finally obtain $\dot{a}^2$ as a function of $a$. It is easy to check that $V(a_0)=0$ and, with the help of Eq. (\ref{CondEstatico}), that also $V'(a_0)=0$. The second derivative of the potential, evaluated at $a_0$ has the form
\begin{equation}
V''(a_0)= A''(a_0)-\frac{20}{a_0^2}A(a_0).
\label{potencial2der}
\end{equation}
Then, we can determine that the configuration with radius $a_0$ is stable under radial perturbations if and only if $V''(a_0)>0$.

\section{Wormholes with charge}\label{charge}

\begin{figure}[t!]
\centering
\includegraphics[width=0.4\textwidth]{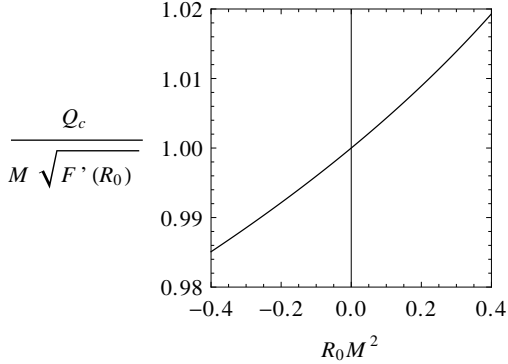}
\caption{Critical charge $Q_c$ in terms of the constant curvature $R_0$ and the mass $M$: if $|Q|< Q_c$ the original manifold presents an inner and an event horizon, which fuse into one when $|Q|= Q_c$; both horizons disappear if $|Q|> Q_c$.}
\label{qcrit}
\end{figure}

Now we analyze an example of application of the formalism introduced in the previous sections. We begin with the action given by 
\begin{equation}
S=\frac{1}{2 \kappa}\int d^4x \sqrt{|g|} (F(R)-\mathcal{F}_{\mu\nu}\mathcal{F}^{\mu\nu}),
\label{action} 
\end{equation} 
where $g=\det (g_{\mu \nu})$,  $\mathcal{F}_{\mu \nu }=\partial _{\mu }\mathcal{A}_{\nu } -\partial _{\nu }\mathcal{A}_{\mu }$ is the electromagnetic tensor, and $F(R)=R+f(R)$ is the function that defines the theory under consideration ($f(R)$ is any suitable function of $R$). The field equations (in the metric formalism) obtained from the action (\ref{action}), considering an electromagnetic potential $\mathcal{A}_{\mu}=(\mathcal{V}(r),0,0,0)$, have the spherically symmetric solution in the form given by Eq. (\ref{metric}), where the metric function \cite{bhfr2} is
\begin{equation} 
A (r) = 1-\frac{2M}{r}+\frac{Q^2}{ F'(R_0) r^2}-\frac{R_0 r^2}{12},
\label{A-metric}
\end{equation}
with $M$ the mass, and $Q$ the charge. This solution has constant curvature $R_0$ and $\mathcal{V}(r)=Q/r$. The value of $R_0$ can be understood in terms of an effective cosmological constant $\Lambda _{\mathrm{eff}}=R_0/4$. It is worth noticing that the squared charge $Q^2$ is corrected by a factor $1/F'(R_0)$ with respect to the General Relativity case. The geometry is singular at $r=0$; the position of the horizons, determined by the zeros of $A (r)$, are given by the positive solutions of a fourth degree polynomial if $R_0 \neq 0$, or of a quadratic function if $R_0=0$. In the case $R_0>0$, for small values of $|Q|$ there are three horizons: the inner $r_i$, the event $r_h$ and the cosmological $r_c$ ones; when the charge is large enough, i.e. $|Q|=Q_c$, the inner and the event horizons fuse into one; finally if $|Q|>Q_c$ there is a naked singularity at the origin and only one horizon in $r_c$. When $R_0\le 0$, for small values of $|Q|$ there are two horizons: the inner $r_i$ and the event $r_h$ ones; when the charge is $|Q|=Q_c$, the inner and the event horizons fuse into one; finally if $|Q|>Q_c$ there is a naked singularity and no horizons. The critical value of the charge $Q_c$ is plotted in Fig. \ref{qcrit}. 

\begin{figure}[t!]
\centering
\includegraphics[width=1.0\textwidth]{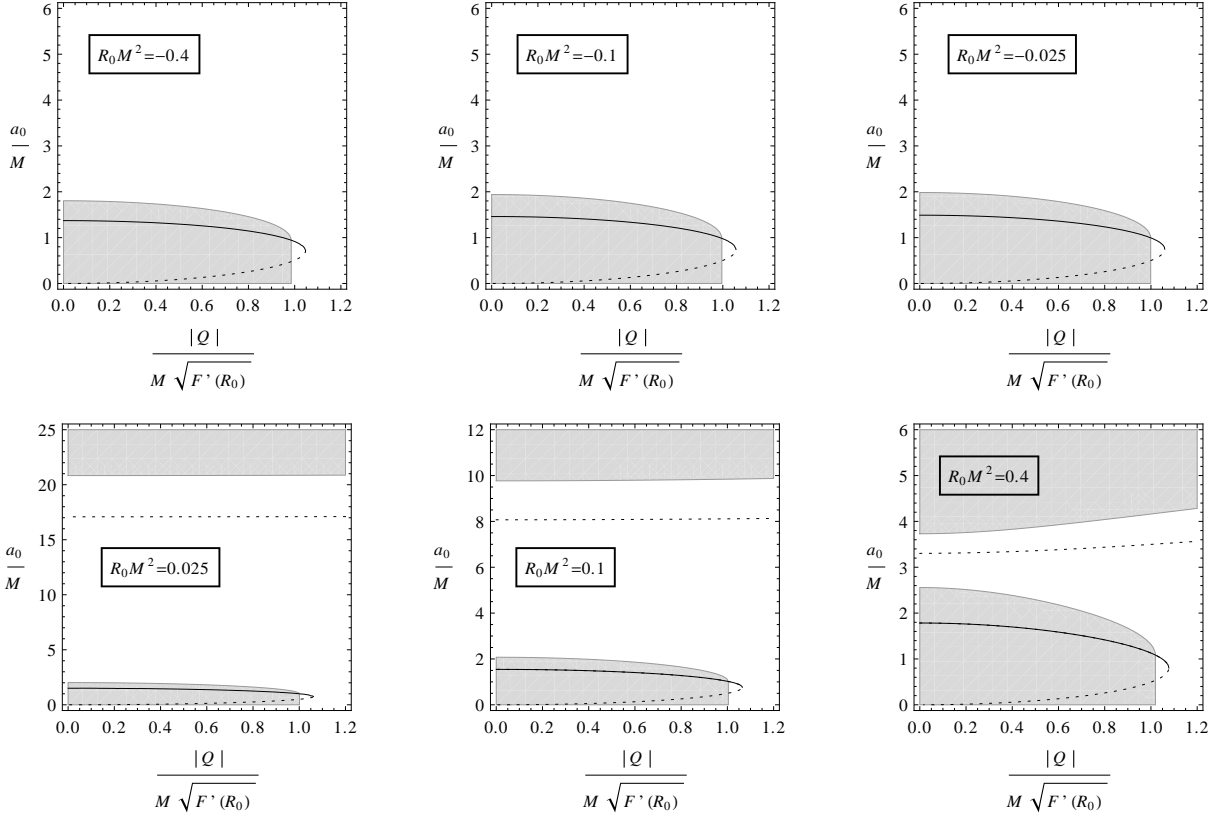}
\caption{Stability of wormholes in $F(R)$ theories for different values of the constant curvature $R_0$. Solid curves represent static stable solutions with throat radius $a_{0}$, while dotted lines correspond to unstable ones; $M$ and $Q$ are the ADM mass and charge, respectively. Gray zones are non--physical. Note that two plots have a different scale in the vertical axis.}
\label{estab}
\end{figure}

In our wormhole construction, the radius $a$ of the throat is taken large enough to avoid the presence of the inner and the event horizons, and the singularity; when corresponds it is also smaller than the cosmological horizon. If a  static solution exists for a given set of parameters, the throat radius  $a_{0}$ should satisfy Eq. (\ref{CondEstatico-bis}), which for the particular metric function given by Eq. (\ref{A-metric}) adopts the form
\begin{equation}
\frac{R_0}{2}a_0^4-4a_0^2+6Ma_0-\frac{2Q^2}{ F'(R_0)}=0.
\label{extra} 
\end{equation}
From Eqs. (\ref{e13}) and (\ref{e14}), the energy density and the pressure in this case are
\begin{equation} 
\sigma_0=\frac{-4 F'(R_0)}{\kappa a_0}\sqrt{1-\frac{2M}{a_0}+\frac{Q^2 }{F'(R_0)a_0^2}-\frac{R_0 a_0^2}{12}},
\label{e13metric}
\end{equation}
and
\begin{equation}
p_0=\frac{-2 F'(R_0)}{\kappa a_0}\sqrt{1-\frac{2M}{a_0}+\frac{Q^2 }{F'(R_0)a_0^2}-\frac{R_0 a_0^2}{12}}.
\label{e14metric}
\end{equation}
Then, if one expects that the term in the metric (\ref{A-metric}) corresponding to the charge has the same sign as in General Relativity, the presence of exotic matter at the throat is required. As mentioned above, the inequality $F'(R_0)>0$ implies a positive effective gravitational constant and the absence of ghosts. So in what follows we assume that $F'(R_0)>0$. Using Eq. (\ref{potencial}), the potential is
\begin{equation}
V(a)= 1-\frac{2M}{a}+\frac{Q^2}{F'(R_0) a^2}-\frac{a^2 R_0}{12}-\frac{a_0^4}{a^4}\left(1-\frac{2M}{a_0}+\frac{Q^2 }{F'(R_0)a_0^2}-\frac{a_0^2 R_0}{12} \right), 
\label{potencialmetric}
\end{equation}
from which we verify that $V(a_0)=0$ and $V'(a_0)=0$, and we obtain that
\begin{equation}
V''(a_0)= \frac{3}{2}R_0-\frac{20}{a_0^2}+\frac{36 M}{a_0^3}-\frac{14 Q^2}{F'(R_0) a_0^4}.
\label{potencial2dmetric}
\end{equation}
As stated above, $V''(a_0)>0$ is the condition for stability under radial perturbations. 

The results are presented graphically in Fig. \ref{estab}, in which we have chosen the most representative figures. The stable solutions are shown with solid lines, while the dotted lines correspond to unstable configurations. The regions shaded in gray have no physical meaning, because they correspond to the zones inside the event horizon or the cosmological horizon of the original manifold which are removed in the construction of the wormholes. The results present important changes around the value of $Q_c/M$, where $Q_{c}$ is the critical charge, corresponding to the value of charge from which the original metric used for the construction loses the inner and the event horizons. The solutions show a behavior which strongly depends on the values of $R_0 M^2$. However, the different values of $F'(R_0)$ affect the results only in the form of an ``effective charge'' $Q/\sqrt{F'(R_0)}$. Depending on the sign of $R_0$, we have
\begin{itemize}
\item For $R_0\le 0$, no static solutions are present if $|Q| \le Q_c$. As the charge grows, i.e. $|Q|>Q_c$, two static solutions appear: one stable and the other unstable. Then, the static solutions fuse into one; finally for larger values of charge they disappear.
\item For $R_0>0$, when $|Q|>Q_c$ there exist two static solutions (one stable and the other unstable) with a similar behavior as in the $R_0<0$ case, but also a third static solution with a larger value of $a_0/M$ is always present. This third solution is unstable for any value of the charge.
\end{itemize}
In any case, we see that stable static solutions are always obtained if appropriate values of mass and charge are taken for a given function $F(R)$ and constant curvature $R_0$.

\section{Conclusions}\label{conclu}

We have constructed a class of spherically symmetric wormholes by using the thin--shell formalism in $F(R)$ theories; the surface that joins the two equal copies of a solution with constant curvature $R_0$ corresponds to the throat. We have shown that the matter at the throat should satisfy the equation of state $p=\sigma /2$. The condition $F'(R_0)>0$, required to have a positive effective gravitational constant and a non--ghost graviton, results in exotic matter at the throat. We also have obtained the condition for the stability of static configurations under perturbations preserving the symmetry. In particular, we have applied the formalism to wormholes with mass $M$ and charge $Q$. As we have assumed that $F'(R_0)>0$, the term associated with the charge has the same sign as in General Relativity, and the matter at the throat is exotic. We have found that stable solutions are possible for appropriate values of the parameters, for both positive and non--positive $R_0$. In the first case, one static solution is always also present, being always unstable for any value of the charge; for large values of $|Q|/ (M\sqrt{F'(R_0)})$ two static solutions are also present within a short range of charge, one of them is stable, while the other is unstable. In the second case, a large value of $|Q|/ (M\sqrt{F'(R_0)})$ is required to have two static solutions for a short range of charge, one of them is stable and the other is unstable.  The qualitative aspects of the results do not depend on the particular form of the function $F(R)$, each theory only manifests itself through the constant $F'(R_0)$, in the form of an effective charge $Q/F'(R_0)$.

\section*{Acknowledgments}

This work has been supported by CONICET and Universidad de Buenos Aires.

\end{document}